\documentclass[twocolumn,aps,floats,superscriptaddress,showpacs]{revtex4}
\usepackage{amsmath}

\usepackage{epsfig}
\usepackage{graphicx}
\usepackage{dcolumn}
\usepackage{bm}

\def\gapp{\lower.35em\hbox{$\stackrel{\textstyle>}{\sim}$}}
\def\lapp{\lower.35em\hbox{$\stackrel{\textstyle<}{\sim}$}}

\begin{document}
\bibliographystyle{apsrev}
%


\title{Charge inhomogeneities due to smooth ripples in graphene sheets}

\author{Fernando de Juan, Alberto Cortijo, and Mar\'{\i}a A. H. Vozmediano}
\affiliation{Unidad Asociada CSIC-UC3M,
Instituto de Ciencia de Materiales de Madrid,\\
CSIC, Cantoblanco, E-28049 Madrid, Spain.}

\date{\today}
\begin{abstract}
We study the effect of the curved ripples observed in the free
standing graphene samples on the electronic structure of the
system. We model the ripples as  smooth curved bumps and compute
the Green's function of the Dirac fermions in the curved surface.
Curved regions modify the Fermi velocity that becomes a function
of the point on the graphene surface and induce energy dependent
oscillations in the local density of states around the position of
the bump. This effect is due to the change of the Pauli matrices
with the position  and is independent of the well known gauge
potential generated by the spin connection. The corrections to the
density of states are only due to the first effect and are
estimated to be of a few percent of the flat density of states.
Local probes such as scanning tunnel microscopy  should be able to
observe the predicted correlation of the morphology with the
electronics. We discuss the connection of the present work with a
recent observation of charge anisotropy in graphene and argue that
the ripples can provide an alternative explanation.

\end{abstract}
%
%
%
%

\maketitle

\section{Introduction}

Recent experiments  of transmission electron  (TEM)
\cite{Metal07,Metal07b}  show that the suspended graphene samples
exhibit an apparently random spontaneous curvature that can be
visualized as ripples of various sizes that can reach a few
angstroms high and several nanometers long. Observed ripples have
also been reported later in scanning tunnelling microscopy
\cite{Setal07,Ietal07} (STM). The issue of curvature of graphene
and its possible influence on the electronic properties have been
addressed before. Inspired by the physics of nanotubes and
fullerenes most of the works done on the electronic properties of
curved graphene dealt with curvature induced by topological
defects \cite{TT94,CR01,GGV01,CV07a,CV07b}. In these works it was
shown that conical singularities in the average flat graphene
sheet induce characteristic charge anisotropies that could be seen
in (STM) or (TEM). Charge anisotropies in monolayer graphene have
been recently observed in scanning single electron transistor
experiments \cite{Y07} that show a distribution of electron-hole
puddles at the Fermi level which could be responsible for the
minimal conductivity of graphene \cite{KNG06}. Charge anisotropies
have also been reported in Electrostatic Force Microscopy
experiments done in graphite \cite{GE06}. Charge inhomogeneities
in graphene have been computed recently associated to impurity
states \cite{WBetal06} or elastic strain \cite{CK07} of a  nature
very similar to the one obtained previously with topological
defects. It is by now clear that fluctuations in structure and
charge are closely related and it seems that most types of
disorder will induce charge anisotropies near the Dirac cone.

In this work we study the effect of smooth curved portions of the
type described in the experiments on the density of states of
graphene by coupling the Dirac equation describing the low energy
electronic properties of graphene to a general metric describing a
smooth curved piece in the average flat surface. The formalism is
the same  used in \cite{CV07a,CV07b} to study the effect of
pentagon and heptagon rings but here there are no curvature
singularities. We obtain that the charge density couples to the
shape of the graphene surface. The local density of states
oscillates with the energy and the maximal correction has a
spacial extent of the order of the defect. The estimated value of
the relative correction at the typical energies explored in the
experiments can be of the order of a few percent of the flat value
much bigger than the one obtained with elasticity models. The
predicted correlation of the morphology of the sample with the
electronics should be observed in local probes as STM or TEM
experiments. The present mechanism can  explain qualitatively the
recent observations of charge inhomogeneities
 at zero energy found in graphene \cite{Y07} as due to the space
variation of the Fermi velocity induced by the curvature.

The article is organized as follows: Section \ref{ham} explores
the consequences of coupling the two dimensional Dirac Hamiltonian
to a curved surface using a smooth gaussian bump as an example. At
this level we see that as a consequence of the curvature two
independent things happen:  the Fermi velocity acquires a non
trivial dependence on the position, and a gauge field is
generated. Section \ref{green} is devoted to the computation of
the local density of states of the system. We first describe the
effects of the curvature in terms of an effective potential and
then compute the electron Green's function in the curved space to
first order in perturbation theory. The perturbative parameter is
related to the deviation of the graphene surface from the flat
plane. Section \ref{results} analyzes the results. In section
\ref{exp} we discuss the possible experimental consequences that
can be extracted from this work in particular in connection with
the anisotropies of ref. \cite{Y07}. In section \ref{final} we
present the conclusions and discussion. The appendix A is devoted
to the definition and computation of the various geometrical
factors associated to the problem.

\section{Two dimensional Dirac equation in a curved space with
polar symmetry} \label{ham}

The most obvious way to study the effects of the topography of the
sample on the electronic degrees of freedom is to couple the Dirac
equation that governs the low energy electronics to the curved
surface. This approach has been applied to study curved fullerenes
\cite{GGV92,GGV93,GGV93A,OK05,KO06} and to compute the response of
electromagnetic charges to conical defects in planar graphene
\cite{FM94}  in the framework of the equivalence between the
theory of defects in solids and the three-dimensional gravity
\cite{DV80,KV92}. The coupling of the electronic degrees of
freedom of planar graphene to conical defects has been explored in
\cite{CV07a,CV07b,CV07d}. There it was found that a distribution
of pentagons and heptagons induces characteristic inhomogeneities
in the density of states of the graphene surface. In order to
investigate the effect of pure curvature on the electronic
properties of graphene, the conical defects studied previously
present two difficulties. First they correspond to surfaces with
zero intrinsic curvature; moreover the extrinsic curvature is
accumulated at the apex of the cone where the surface has a
singularity. It is then not clear if the results obtained -- which
look similar to the ones got with vacancies in \cite{WBetal06} --
are due to the singularity or to the curvature. To disentangle the
two effects we have studied the density of states of a flat
graphene sheet with a  smooth curved portion with intrinsic
curvature.

The  massless Dirac equation in a curved spacetime is given by:
\begin{equation}
i\gamma^\mu(\partial_\mu+\Omega_\mu)\Psi=0, \label{diracc}
\end{equation}
where $\gamma^\mu=(\gamma^0, v_F \gamma^i)$, i=1,2.  These  curved
$\gamma$ matrices satisfy the anticommutation relations
\begin{equation}
\{\gamma^{\mu},\gamma^{\nu}\}=2g^{\mu\nu}(x),
\end{equation}
and in general become functions of the point in spacetime $x =(t,
{\bf r})$. $\Omega_{\mu}(x)$ is the spin connection of the spinor
field that can be calculated using the tetrad formalism
\cite{BD82} and will be defined in the appendix A.

What is needed is a metric describing the curvature of the
surface. We will study the general case of a smooth protuberance
fitting without singularities  in the average flat graphene sheet.
We start by embedding a two-dimensional surface with polar
symmetry in three-dimensional space (described in cylindrical
coordinates). The surface is defined by a function $z(r)$ giving
the height with respect to the flat surface z=0, and parametrized
by the polar coordinates of its projection onto the z=0 plane. The
metric for this surface is obtained as follows: We compute
\begin{equation}
dz^{2}=(\frac{dz}{dr})^{2}dr^{2}\equiv \alpha f(r)dr^{2},
\label{surface}
\end{equation}
 and substitute for the line element:
\begin{equation}
ds^{2}=dr^{2}+r^{2}d\theta^{2}+dz^{2}=\left(1+\alpha
f(r)\right)dr^{2}+r^{2}d\theta^{2}. \label{generalmetric}
\end{equation}
 We will assume that the surface is asymptotically flat at long
distances, so that $f$ decays with $r$ fast enough. We will also
require $f$ to go to zero sufficiently fast in $r=0$ so that the
surface is smooth there.
\begin{figure}
  \begin{center}
    \epsfig{file=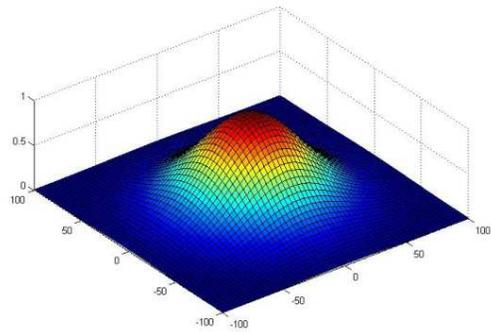,height=5cm}
    \caption{(Color online) The smooth curved bump discussed in the text. }
    \label{dibujo}
\end{center}
\end{figure}
\begin{figure}
  \begin{center}
    \epsfig{file=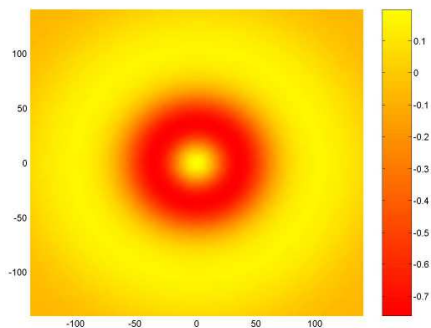,height=5cm}
    \caption{(Color online) Effect of the curved bump of fig. \ref{dibujo}
    on the local density of states of the graphene sheet. }
    \label{polares}
\end{center}
\end{figure}
For clarity, we work out as an example the gaussian bump shown in
Fig. \ref{dibujo} defined by:
\begin{equation}
z=A\exp(-r^{2}/b^{2}),
\label{gaussianformula}
\end{equation}
so that
\begin{equation}
dz^{2}=\frac{A^{2}}{b^{4}}4r^{2}\exp(-2r^{2}/b^{2})dr^{2},
\end{equation}
which corresponds to eq. (\ref{generalmetric}) with
$$\alpha=(A/b)^{2}\;,\; f(r)=4(r/b)^{2}\exp(-2r^{2}/b^{2}).$$
The ratio of the height to the mean width of the gaussian will be
our perturbative parameter. Values of $b$ of the order of 0.1 to
0.3 times the height $A$ give bumps of magnitude and shape
comparable with the ripples reported in \cite{Metal06,Metal07}.

Since we are going to work with the Dirac equation, our
calculation is formally relativistic and the full space-time
metric must be used. The line element reads:
\begin{equation}
ds^{2}=dt^{2}-\left(1+\alpha f\right)dr^{2}-r^{2}d\theta^{2},
\label{fullmetric}
\end{equation}
from where we can write  the metric in a more usual form:
\begin{equation}
g_{\mu\nu}=
\left(%
\begin{array}{ccc}
  1 & 0 & 0 \\
  0 & -(1+\alpha f(r)) & 0 \\
  0 & 0 & -r^2 \\
\end{array}%
\right). \label{metric}
\end{equation}
Since we are dealing with a problem with axial symmetry, we will
work in polar coordinates. The Dirac Hamiltonian of the plane
(flat) in polar coordinates can be written as
\begin{equation}
H_{flat}=\hbar v_F
\left(%
\begin{array}{cc}
  0 & \partial_r+i\frac{\partial_\theta}{r}+\frac{1}{2r} \\
  \partial_r-i\frac{\partial_\theta}{r}+\frac{1}{2r} & 0  \\
\end{array}%
\right), \label{Hflat}
\end{equation}
where, as discussed in Appendix A,  a constant spin connection has
been generated that is a "pure gauge" and can be rotated away by a
different choice of local coordinates\footnote{We can see that the
effective magnetic field generated by this potential is zero.}.
The calculation of the different geometric factors for the metric
(\ref{fullmetric}) is given in the appendix A. Adding them
together we get the curved Hamiltonian
\begin{widetext}
\begin{equation}
H_{curved}=\hbar v_F
\left(%
\begin{array}{cc}
  0 & (1+\alpha f(r))^{-1/2}\partial_r+i\frac{\partial_\theta}{r}+A_\theta \\
  (1+\alpha f(r))^{-1/2}\partial_r-i\frac{\partial_\theta}{r}+A_\theta & 0  \\
\end{array}
\right), \label{Hcurved}
\end{equation}
\end{widetext}
where the effective gauge potential is related to the coefficient
of the spin connection computed in the appendix (\ref{spincon}) by
\begin{equation}
A_\theta=\frac{\Omega_\theta}{2r}=\frac{1-(1+\alpha
f)^{-1/2}}{2r}.
\end{equation}
Comparing  (\ref{Hflat}) with (\ref{Hcurved}) we can read that the
curved bump has produced an effective Fermi velocity
$\tilde{v}_{r}$ in the radial direction given by
\begin{equation}
\tilde{v}_r(r,\theta)=v_F(1+\alpha f(r))^{-1/2},
\end{equation}
and an effective magnetic field perpendicular to the graphene
sheet given by
\begin{equation}
B_z=-\frac{1}{r}\partial_r(rA_\theta)=\frac{1}{4r}\frac{\alpha
f'}{(1+\alpha f)^{3/2}}. \label{B}
\end{equation}
The magnitude of this effective magnetic field is estimated to be
of the order of 0.5 to 2-3 Tesla in the region spanned by the
bump, compatible with the estimations given in \cite{Metal06}, and
it will play the same  role  in the issue of the weak localization
of graphene  as the effective magnetic  fields discussed there and
in \cite{MG06}.

We note that in general the effective Fermi velocity will be
smaller in magnitude than the free one. For a general curved
surface described in polar coordinates by $z=z(r)$, the effective
Fermi velocity will be
\begin{equation}
v_r=\frac{v_0}{\sqrt{1+z'(r)^2}}.
\end{equation}
In a most general case we will have the two components of the
velocity changed but always to a smaller value.

 In the next section we specify
the method to compute the density of states through the electron
Green's function. For this purpose it is more convenient to follow
the appendix and rewrite the Dirac equation in the form
\begin{equation}
\left[i\gamma^{0}\partial_{0}+i\Gamma(\theta)\partial_{r}+
i\Gamma'(\theta)\frac{\partial_{\theta}}{r}+V(r,\theta)\right]\Psi=0,
\end{equation}
which is the flat Dirac equation in polar coordinates in a sort of
potential V given by:
\begin{equation}\label{potential}
V(r,\theta)=i\Gamma(\theta)\left[1-(1+\alpha
f)^{-1/2}\right](\frac{1}{2r}-\partial_{r}).
\end{equation}
where $\Gamma(\theta)=\gamma^1\cos\theta+\gamma^2\sin\theta$. This
effective potential will be used in the next section to compute
the local density of states of the system. We can read in it  two
different terms related with our previous discussion: the
derivative term has its origin in the effective r-dependent Fermi
velocity and the term proportional to $1/r$ comes from the
effective gauge field.

\section{The Green's function in a curved spacetime. Approximations}
\label{green}

 We drop the polar coordinates for a moment to
outline the procedure to obtain the Dirac propagator in a sort of
perturbative expansion.

Since we will be interested in small deformations from the flat
membrane, we will compute the first order correction to the
``flat'' propagator in the small parameter $\alpha$. In our
example,  $\alpha=(A/b)^{2}$ is the (squared) height to length
ratio of the gaussian, so for typical ripples in graphene
$\alpha\approx 0.01$, since this ratio is of the order of 0.1
\cite{Metal07}.

The equation for the exact propagator in the curved space-time is:
\begin{equation}
i\gamma^{\alpha}e_{\alpha}^{\
\mu}\left(\partial_{\mu}+\Omega_{\mu}\right)\
G(x,x')=\delta(x-x')(-g)^{-\frac{1}{2}},
\end{equation}
where $\Omega_\mu$ is the spin connection computed in the appendix
(\ref{spincon}) and $\sqrt{-g}$ is the determinant of the metric
given in (\ref{jacobian}). Since the flat propagator equation is
recovered when $\alpha=0$, expanding the left hand side to first
order in $\alpha$ will give this flat equation plus a first order
general term that we will call V. We can as well expand
$(-g)^{-1/2}=1-\alpha f(x)$, and sending the f term to the left
hand side we get an equation resembling the flat propagator
equation in a sort of potential generated by the metric.
\begin{equation}
(i\gamma^{\mu}\partial_{\mu}+V)G(x,x')+\alpha f(x) \delta(x-x')=
\delta(x-x').
\end{equation}

This equation can be solved by the usual pertubative expansion of
G in a potential. Note that two approximations to order $\alpha$
are taking place: first, we expand the curved space exact equation
for G to make it resemble the flat equation in a potential, and then
we use a perturbative expansion in the potential to get the first order correction to G.

The solution for G is given by:
\begin{eqnarray}\label{guno}
G(x,x')=G_{0}(x-x')-G_{0}(x-x')\alpha f(x)+ \nonumber \\
\int dx''G_{0}(x,x'')V(x'')G_{0}(x'',x').
\end{eqnarray}

We now proceed to use this expansion in our particular case. The determinant of the metric is just:
\begin{equation}
g^{-1/2}=\frac{(1+\alpha f)^{-1/2}}{r}\approx \frac{1-\frac{\alpha
f}{2}}{r}.
\end{equation}
Expanding (\ref{potential}) to first order in $\alpha$ we get:
\begin{equation}\label{potential1}
V(r,\theta)=i\Gamma(\theta)\left[\frac{1}{2}\alpha
f(r)\right](\frac{1}{2r}-\partial_{r})
\end{equation}
Noting that the flat $\delta$ function in polar coordinates is
$\delta(r-r')/r$, we can use eq. (\ref{guno}) in polar coordinates
to get the first order correction to the propagator.

Next we need the Dirac propagator in polar coordinates. We will
get it by noting that the Dirac and Klein-Gordon propagators are
related by:
\begin{equation}\label{DKG}
G^{0}_{D}=D(x)G^{0}_{KG},
\end{equation}
where $D(x)$ is the Dirac operator.
The Dirac operator in polar coordinates is:
\begin{equation}\label{diracpolar}
\gamma^{0}E+i\left[\Gamma(\theta)\partial_{r}+\Gamma'(\theta)\frac{\partial_{\theta}}{r}\right],
\end{equation}
where we have defined:
\begin{eqnarray}
\Gamma(\theta)=\gamma^{1}\cos(\theta)+\gamma^{2}\sin(\theta)\;,
&&\nonumber  \\
\Gamma'(\theta)=-\gamma^{1}\sin(\theta)+\gamma^{2}\cos(\theta).
\end{eqnarray}
The Klein-Gordon propagator in polar coordinates is
\begin{equation}
G_{KG}(r-r',E)=\frac{-i}{4}H_{0}(E\left|r-r'\right|),
\end{equation}
where $H_0$ is the zeroth order Hankel function \cite{GR80}.
Applying eqs. (\ref{diracpolar}) and  (\ref{DKG}) we get:
\begin{eqnarray}\label{diracprop}
G_{D}(r-r',E)=\frac{-iE}{4}\gamma^{0}H_{0}(E\left|r-r'\right|)-
\nonumber \\
 \frac{E}{4\left|r-r'\right|}H_{1}(E\left|r-r'\right|)
\left[r\Gamma(\theta)-r'\Gamma(\theta')\right].
\end{eqnarray}
Now we just use eqs. (\ref{guno}) and (\ref{diracprop}), and the
first order potential (\ref{potential1}) to get the local density
of states
\begin{equation}
\rho(E,\textbf{r})=-\frac{1}{\pi}Im Tr
[G(E,\textbf{r},\textbf{r})\gamma^{0}].
\end{equation}
After two partial integrations we obtain the local density of
states as:
\begin{widetext}
\begin{eqnarray}
\rho(r',E)=g\frac{E}{2\pi}\frac{1}{(v_{F}\hbar)^{2}}[1-\frac{\alpha
f(r')}{2}+
 \frac{\alpha}{8}\int dr d\theta .\qquad\qquad\qquad\qquad\nonumber\\
 \left[4E^{2}f(r)r\frac{r-r'\cos(\theta-\theta')}
{(r^{2}+r'^{2}-2rr'\cos(\theta-\theta'))^{1/2}}Y_{1}(E\Delta r)
J_{1}(E\Delta r)-
\left[2f'(r)+r f''(r)\right]Y_{0}(E\Delta r)J_{0}(E\Delta r)
\right] ] , \label{LDOS}
\end{eqnarray}
\end{widetext}
where g is the spin and valley degeneracy, $\Delta(r)\equiv\vert
r-r'\vert$ and $Y_i, J_i$ are Bessel functions \cite{GR80}. We
will analyze this result in the next section.

\section{Results}
\label{results}
\begin{figure}
  \begin{center}
    \epsfig{file=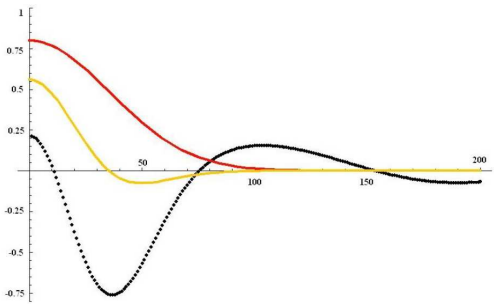,height=4cm}
    \caption{(Color online) Correction to the density of states
    (dotted line) in arbitrary units, versus the
    shape of defect (red -- upper -- line) and curvature of the defect (yellow --middle-- line)
    for a gaussian bump of an average width of 50 $\AA$.    }
    \label{comp}
\end{center}
\end{figure}
The main result of this work is the confirmation of the fact that
the morphology of the graphene samples is correlated with the
electronic properties. In particular the presence of ripples,
irrespective of their origin, induces corrections to the density
of states and will affect the transport properties.  The analysis
of this work is quite general and the results will apply to any
smooth surface. We have considered as an example a gaussian bump
of an average spacial extent of 2 to 20 nanometers similar to the
ripples described in \cite{Metal06,Metal07}. As can be read from
eq. (\ref{LDOS}) due to the Bessel functions the corrections to
the local density of states (LDOS) show spacial oscillations whose
frequency grows with the energy and whose amplitude  decays with
the distance as (1/r). Fig. \ref{polares} shows the correction to
the LDOS induced by the shape of Fig. \ref{dibujo} with a mean
width $b$ of $50 \AA$  and for an energy of $E=0.1$ eV.  The color
scale is indicated in the figure. Lighter (darker) color means a
positive (negative) contribution with respect to the flat graphene
sheet at the given energy. The maximal value of the correction
related to the bare LDOS $\rho(E, r)/\rho_0(E,r)$ for a bump of a
ratio $A/b\sim 0.1$ is of the order of a 1 percent. If the height
of the ripples goes up to 0.3, the maximal value of the correction
to the LDOS due to the curved portion can reach a 10 percent.
Larger values are possible but they would correspond to higher
values of our perturbative parameter $\alpha$ and made the result
questionable. The bare LDOS being proportional to the energy (E),
allows to relate the correction obtained with the energy at which
the experiment takes place.

At low energies the maximal absolute value of the correction is
correlated with the zero of the curvature which, in the particular
case of the gaussian bump, coincides with the mean width
$\sqrt{2}b$. Fig. \ref{comp} shows the correction to the local
density of states (dotted line) in arbitrary units, versus the
shape of defect (red -- upper -- line) and the curvature of the
defect (yellow --middle-- line) for a gaussian bump of an average
width of 50 $\AA$. The energy is 0.1 eV. The figure represents
real space in polar coordinates. The horizontal axis is the r
coordinate while the vertical axis represents real height in the
case of the upper line (shape of the gaussian bump considered).
The middle line (curvature) given in eq. (\ref{curvature})  is not
measured in units of length. Finally the lowest curve gives the
correction to the LDOS in arbitrary units. This results seems to
be at odd with related works based on topological defects
\cite{Metal06,MG06} or elasticity \cite{CK07} that correlate the
physical effect of curved portions with the actual value of the
geometrical curvature.

We notice here that of the two effects of the curvature discussed
in this work, only the effective magnetic field coming from the
spin connection can be compared with previous works. In our case
and in the general situation of having a smooth shape with axial
symmetry, the effect of the effective gauge field vanishes at
first order in perturbation theory and the corrections to the
local density of states come exclusively from the spacial
dependence of the Fermi velocity. This effect has not been noticed
before because most of the previous works coupling the Dirac
equation to curved space dealt either with spherical shapes where
the correction to the Fermi velocity is constant and can be scaled
out, or with conical shapes whose intrinsic curvature is
accumulated at the apex singularity.

A discussion on the various gauge fields that arise in the physics
of graphene and their physical consequences will be published
elsewhere \cite{CJV07}.

 For simplicity we have modelled the ripples with shapes that
are axially symmetric and the axial symmetry is explicit in the
results. More general shapes would made the calculation much more
complicated without altering the main results.

We can observe in figs. \ref{polares} and \ref{comp} that the
maximal value of the correction to the LDOS is concentrated in the
region spanned by  the bump. In an STM experiment done on the
curved surface the results on the LDOS plotted on the flat surface
should resemble the rings of fig. \ref{polares} and would be in a
very precise correspondence with the morphology of the sample.

\begin{figure}
  \begin{center}
    \epsfig{file=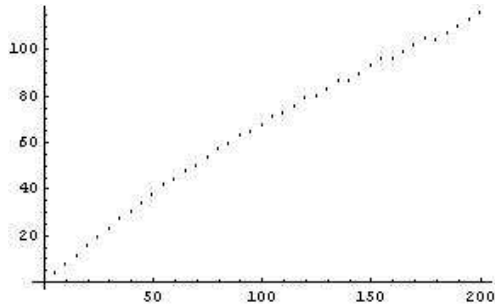,height=4cm}
    \caption{Plot of the location of the maximal correction of the LDOS for a fixed energy as a
    function of the extension of the Gaussian b.
    }
    \label{minconb}
\end{center}
\end{figure}

To verify the apparent pining of the maximal value of the
correction to the zero of the curvature observed in \ref{comp} we
have explored a set of gaussian shapes with different widths. The
result is plotted  in fig. \ref{minconb}. The correlation is good
for the physical values of the bump from 2 to 5 nanometers ($50
\AA$).
\begin{figure}
  \begin{center}
    \epsfig{file=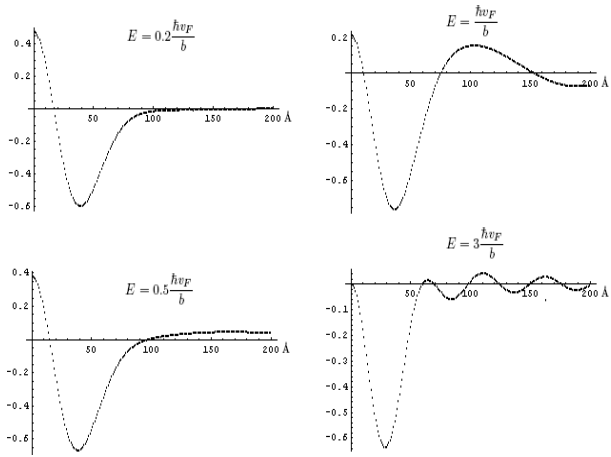,height=6cm,width=8cm}
    \caption{Correction to the density of states of
    a bump of fixed width (50 \AA) for different values of the energy. }
    \label{6energies}
\end{center}
\end{figure}
Fig. \ref{6energies} shows the dependence of the correction to the
LDOS with the energy. The four plots show the correction induced
by the same bump of a width of 5 nanometers  for growing values of
the energy.

\section{Possible experimental consequences. }
\label{exp}

 We will here discuss the possible connection of this
work with the recent experimental observations of electron-hole
puddles of ref. \cite{Y07}\footnote{We thank A. Geim for
suggesting us to explore in this direction.}.

The main experimental facts that we have to address are the
intrinsic disorder length scale of approximately 30 nm and the
magnitude of the density fluctuations of an average value $\Delta
n\sim\pm 4 . 10^{10} cm^{-2}$ that can reach a maximum value of
$10^{11} cm^{-2}$.



In the experiment described in \cite{Y07}, the local inverse
compressibility $\partial \mu /\partial n$ is measured as a
function of a back-gate voltage. In the case of having a flat
graphene sheet this quantity can be computed analytically using
the  free dispersion relation and density of states, and it
becomes more complicated when the latter is corrected due to the
curvature.

Before entering in a more detailed analysis we can explain
qualitatively the observation of \cite{Y07} as follows: From our
result of Section \ref{green} we note that a variable correction
to the density of states of the type shown in (\ref{LDOS}) causes
a global, constant variation of the Dirac point $E_{D}$ with
respect to the charge neutrality point, because of the
redistribution of the electronic charge when the density of states
is modified locally (notice from Fig. \ref{comp} that the
corrected LDOS is always lower than the flat one). This mechanism
is the one proposed in ref. \cite{CK07} in a different context. An
observation like the one done in \cite{Y07} will see electron and
hole puddles of the size of the ripples without the need to argue
for charged impurities or other kinds of disorder. To give a
number, we can estimate the density of carriers that we would need
to induce to reach the Dirac point at a given position of the
sample in this case as the integral of the correction of the
density of states from some lower cutoff (of the order of  -0.7 eV
) to 0. In the case of the maximal correction we find this density
to be of order $\Delta n\sim\pm 10^{11} cm^{-2}$.

For a more quantitative discussion we consider a flat graphene
sheet with some general $E_{D}$ measured with respect to the
charge neutrality point $E_{cnp}$, which we set to zero. The local
density of states would be given in such a case by:
\begin{equation}\label{ldosflat}
\rho (E,r)=\frac{\left|E-E_{D}(r)\right|}{2\pi}\frac{g_{S}}{(\hbar
v_{F})^{2}},
\end{equation}
where $g_{S}$=4 is the degeneracy (spin and valley). When a
voltage is applied, a density of carriers is induced locally, and
thus the local chemical potential varies as well, though not
linearly with the voltage. The induced number of carriers in this
case can be computed as:

\begin{equation}
n=\int_{0}^{\mu}dE\frac{\left|E-E_{D}\right|}{2\pi}\frac{g_{S}}{(\hbar
v_{F})^{2}}=
\int_{-E_{D}}^{-E_{D}+\mu}dE\frac{\left|E\right|}{2\pi}\frac{g_{S}}{(\hbar
v_{F})^{2}}.
\end{equation}

The integration yields, taking appropriate care of the signs due
to the absolute value:

\begin{equation}\label{n}
n=\frac{\left[sign(-E_{D}+\mu)(-E_{D}+\mu)^{2}-sign(-E_{D})E_{D}^{2}\right]
g_{S}}{4\pi(\hbar v_{F})^{2}}.
\end{equation}

The rate of change $\partial n / \partial \mu$ is again
$\rho(-E_{D}+\mu)$, and solving for $\mu$ in (\ref{n}),
substituting the value and inverting, we obtain:
\begin{equation}\label{componenda}
\frac{\partial \mu}{\partial n}=\frac{ \hbar v_{F}}{\left| \frac{n
g_{S}}{\pi}-\frac{E_{D}^{2}sign(E_{D})g_{S}^{2}}{4\pi^{2}(\hbar
v_{F})^{2}}\right|^{1/2}},
\end{equation}
which reduces to the formula given in ref. \cite{Y07} when
$E_{D}=0$. We can see that if $E_{D}$ is shifted from zero, the
curve for the inverse compressibility keeps its shape but it is
displaced in the $n$ axis, as shown in the experiment.

However there is an alternative interpretation of the displacement
of the curves. Instead of having a variable $E_{cnp}$ from point
to point, we could explain the shift by having a variable $v_{F}$
as predicted in this work. We consider now the curved case, in
which in a qualitative first approximation we just use eq.
(\ref{componenda}) where $v_{F}^{0}\rightarrow v_{F}(r)$. We can
see in eq. (\ref{componenda}) that this substitution has two
effects: first, it makes $\partial \mu /\partial n$ to scale with
$v_{F}(r)$, and also the divergence is displaced by an amount
proportional to $(E_{D}/v_{F}(r))^{2}$, different from the flat
case. The scaling effect provides an experimental test of the
variable Fermi velocity. Extracting the value of   $v_{F}(r)$ from
the shape of the curves at each point we can get a corrected map
for $E_{D}$. We have analyzed carefully the experimental data of
ref. \cite{Y07}\footnote{We thank A. Yacobi and J. Martin for
kindly providing their data and explaining them to us.} to check
this issue but the precision is not enough to deduce the change in
the slope of the curves from point to point. Perhaps this effect
can be tested in angle resolved photoemission experiments.

A related comment concerns the spatial variation of the Fermi
velocity obtained in this work. Different ripple sizes randomly
distributed over the sample will give rise to a landscape of Fermi
velocities that will directly account for the size of the
inhomogeneities reported.  We note that the effective Fermi
velocity is the (only) fitting parameter used in \cite{Y07}.

\section{Conclusions and discussion}
\label{final} We have studied the effect of curvature  on the
electronic properties of a graphene sheet as due to the spinor
nature of the electron wave function.  With complete generality we
have seen that the curvature of the graphene sheet has two
distinct effects with physical consequences. First the covariant
derivative induces an effective magnetic field (the spin
connection) that depends on the shape of the curved surface. This
effect has been discussed before in the literature and can be
understood in many ways. The second effect is even more
interesting and comes from the curved Pauli matrices. They modify
the Fermi velocity that becomes a function of the point on the
graphene surface and is always lower than the free velocity. This
is not to be confused with the renormalization of the Fermi
velocity induced by electron-electron interactions \cite{GGV94} or
disorder \cite{GGV01,SGV05} where the Fermi velocity becomes
effectively a function of the energy. This effect has not been
noticed before because most of the previous works coupling the
Dirac equation to curved space dealt either with spherical shapes
where the correction to the Fermi velocity is constant and can be
scaled out, or with conical shapes whose intrinsic curvature is
accumulated at the apex singularity.

To model the ripples observed in graphene we have applied the
formalism to study a smooth surface with local curvature that is
asymptotically flat. We have studied the changes in the density of
states by computing the electronic Green's function in the
presence of the curved bump. The effective potential generated by
the curvature induces oscillations in the local density of states
that affect significantly the region spanned by the bump. The
potential couples to the electronic Dirac equation as a gauge
field with two parts. One can be identified with an effective non
abelian magnetic field of the type described previously associated
to topological defects or external charges. This part does not
contribute to the LDOS to first order in or perturbative expansion
due to the traces over the gamma matrices. It will nevertheless
affect other observables and will generate a density of states in
non-perturbative calculations of the type recently presented in
\cite{K07b}. The second piece of the effective potential
represents a gauge field with a derivative coupling and comes from
the modified Fermi velocity. This is the one producing the results
presented in this work.  A conservative estimate gives a relative
correction to the flat density of the order of 0.5 to a few per
cent at the energies usually explored with STM probes that should
be able to correlate the morphology with the predicted correction.
The estimated corrections due to curvature are considerable larger
than the ones recently obtained with the theory of elasticity
\cite{CK07}. They are also different: in most cases due to the
traces over the gamma matrices, the correction to the density of
states that would come from the spin connection vanishes and what
remains arises exclusively from the derivative term in the
potential (the correction to the Fermi velocity). The space
variation of the Fermi velocity induced by the curvature provides
an alternative explanation for the charge anisotropies observed in
\cite{Y07} and reinforces the effect. We propose a mechanism to
disentangle the two effects (a space dependent chemical potential
versus a space dependent Fermi velocity) that would provide an
experimental test of the curvature effects described in this work.

The mechanism proposed based on the variable Fermi velocity due to
the curvature of the sheets  can be distinguished from other
mechanisms by the fact that a variable $v_{F}$ in addition to
displace the position of the curves would change their slope  at
different positions of the sample  providing an experimental test
of the model discussed in this work.

 From the discussion done in the paper is clear
that a part of the described modification of the Fermi velocity
can be transferred by a gauge transformation to the vector
potential and viceversa. Any formulation will produce the same
results when computing observable quantities. In this respect we
should not identify the effective Fermi velocity with any tight
binding parameter. As happens in general relativity, curvature
means interactions. The Fermi velocity measured in the
photoemission experiments or by any other means is the result of
the bare value and the interactions so there is no paradox here.

The presence of a random distribution of curved portions as the
one discussed in this work will affect the transport properties of
the system. In this case the effective magnetic field will play a
role similar to the one discussed in the literature \cite{KY03}.
This issue is currently under investigation and it will be
reported in a different work.

\section{Acknowledgments}
We thank Andre Geim for a careful reading of the manuscript and
for making many useful suggestions. Amir Yacoby  and Jens Martin
for sharing their experimental data with us, to Jens Martin
specially for his patience in providing details of their analysis.
We also thank  J. J. Palacios and F. Guinea for useful
conversations. Support by MEC (Spain) through grant
FIS2005-05478-C02-01 and by the European Union Contract 12881
(NEST) is acknowledged.

\appendix

\section{The gaussian bump. Geometric factors.}
The behavior of  spinors in curved spaces is more complicated than
that of scalar or vector fields because their Lorentz
transformation rules do not generalize easily to arbitrary
coordinate systems. Instead of the usual metric $g_{\mu\nu}$ we
must introduce at each point $X$ described in arbitrary
coordinates, a set of locally inertial coordinates $\xi_X^a$ and
the fielbein fields  $e_\mu^a(x)$, a set of orthonormal vectors
labelled by $a$ that fixes the transformation between the local
and the general coordinates:
\begin{equation}
e_\mu^a(X)\equiv\frac{\partial \xi_X^a(x)}{\partial
x^\mu}\vert_{x=X}.
\end{equation}
We will later compute the fielbein for our particular metric.
 The curved space gamma matrices
$\gamma^\mu(x)$ satisfying the commutation relations
\begin{equation}
\{\gamma_\mu\gamma_\nu\}=2g_{\mu\nu},
\end{equation}
 are related with the
constant, flat space matrices $\gamma^a$ by
\begin{equation}
\gamma^\mu(x)=e^\mu_a\gamma_a.
\end{equation}
 The spin connection $\Omega_\mu(x)$
is defined from the fielbein by
\begin{equation}
\Omega_\mu(x)=\frac{1}{4}\gamma_a\gamma_b
e^a_\lambda(x)g^{\lambda\sigma}(x) \nabla_\mu e^b_\sigma(x),
\end{equation}
 with
\begin{equation}
\nabla_\mu e^a_\sigma=\partial_\mu
e_\sigma^a-\Gamma_{\mu\sigma}^\lambda e_\lambda^a
\end{equation}
 where
$\Gamma_{\mu\sigma}^\lambda$ is the usual affine connection which
is related to the metric tensor by \cite{W72}
\begin{equation}
\Gamma_{\mu\sigma}^\lambda=\frac{1}{2}g^{\nu\lambda}\{\frac{\partial
g_{\sigma\nu}}{\partial x^\mu}+\frac{\partial g_{\mu\nu}}{\partial
x^\sigma}-\frac{\partial g_{\mu\sigma}}{\partial x^\nu}\}.
\label{christoffel}
\end{equation}

 Finally, the determinant of the metric needed to
define a scalar density lagrangian is given by
\begin{equation}
\sqrt{-g}=[\det(g_{\mu\nu})]^{1/2}\;=\;\det[e_\mu^a(x)].
\label{jacobian}
\end{equation}
Before going to the computation  of the geometric factors related
with the metric of eq. (\ref{metric}) we will apply the formalism
to  the flat space in polar coordinates what will help to clarify
the physical discussion later. The two dimensional metric of the
flat space in polar coordinates is
\begin{equation}
g_{\mu\nu}=
\left(%
\begin{array}{cc}
  1 & 0 \\
  0  & r^2 \\
\end{array}%
\right). \label{polarflatmetric}
\end{equation}
The affine connection $\Gamma^{\lambda}_{\mu\nu}$ that only
depends on the metric is
\begin{eqnarray}
\Gamma^{r}_{rr}=0, & \Gamma^{r}_{\theta\theta}= -r , &
\Gamma^{\theta}_{r\theta}=\frac{1}{r} \;\;.
\label{flatconnections}
\end{eqnarray}
Despite the fact that the spin connection appears to be non
trivial, the Riemann curvature which is the "observable" quantity
and does not depend on the choice of coordinates is zero as
corresponds to flat space.

 The fielbein fields $e^{a}_{\; \mu}$ satisfy:
\begin{equation}
\label{flatfielbein1} g_{\mu\nu}=e^a_{\; \mu}e^b_{\;
\nu}\eta_{ab},
\end{equation}
where
$\eta_{ab}$ is the identity matrix in two dimensions.

 This relation does not fix $e^a_{\; \mu}$ uniquely.
There are two  natural choices: one is
\begin{equation}
e_\mu^a=(e^\mu_a)^{-1}=
\left(%
\begin{array}{cc}
1& 0 \\
 0 & r \\
\end{array}%
\right) \label{flat1}
\end{equation}
and the other one is
\begin{equation}
e_\mu^a=
\left(%
\begin{array}{cc}
   \cos\theta & -r\sin\theta \\
   \sin\theta & r\cos\theta \\
\end{array}%
\right). \label{flat2}
\end{equation}
The two choices can be visualized as associated to  flat local
frames that at points of constant r have fixed directions (last)
or rotate with the polar angle (first). The first choice leaves
the gamma matrices as in the cartesian plane and induces a
constant gauge connection whose "associated magnetic field" is
obviously zero.

The second choice transforms the flat gamma matrices
 and does not induce a gauge connection.

Let us now compute the geometric factors related with the metric
of eq. (\ref{metric}):
 The affine connection $\Gamma^{\lambda}_{\mu\nu}$ for the
metric (\ref{metric}) is:
\begin{eqnarray}
\Gamma^{r}_{rr}=\frac{\alpha f'}{2(1+\alpha f)}, &
\Gamma^{r}_{\theta\theta}= - \frac{r}{1+\alpha f}, &
\Gamma^{\theta}_{r\theta}=\frac{1}{r} \;\;, \label{connections}
\end{eqnarray}
where $f'=df/dr$, and the rest of the elements are zero or related
by symmetry.

The geometrical (gaussian) curvature $K$ of the shape given by eq.
 (\ref{gaussianformula})  is
 \begin{equation}
 K(r)=\frac{\alpha f'(r)}{2r(1+\alpha f(r))^2  }.
 \label{curvature}
 \end{equation}

The fielbein fields $e^{a}_{\; \mu}$ satisfy:
\begin{equation}\label{drei}
g_{\mu\nu}=e^a_{\; \mu}e^b_{\; \nu}\eta_{ab},
\end{equation}
where $g_{\mu\nu}$ is our metric given in eq. (\ref{metric}) and
\begin{equation}
\eta_{ab}=
\left(%
\begin{array}{ccc}
  1 & 0 & 0 \\
  0 & -1& 0 \\
  0 & 0 & -1 \\
\end{array}%
\right).
\end{equation}
 We choose the $e^a_{\; \mu}$ to be
\begin{eqnarray}
e^{0}_{\;t}=1 & e^{0}_{\; r}=0 & e^{0}_{\; \theta}=0 \nonumber\\
e^{1}_{\;t}=0 & e^{1}_{\; r}=(1+\alpha f)^{1/2}\cos \theta & e^{1}_{\; \theta}=-r\sin \theta  \nonumber\\
e^{2}_{\;t}=0 & e^{2}_{\; r}=(1+\alpha f)^{1/2}\sin \theta &
e^{2}_{\; \theta}=r \cos \theta;\nonumber\\
\end{eqnarray}

that reduce to the flat set (\ref{flat2}) when $\alpha=0$. Now we
can compute the spin connection coefficients,
\begin{equation}
\omega_{\mu}^{\; ab}=e^a_{\; \nu} \left(\partial
_{\mu} + \Gamma^{\nu}_{\mu\lambda} \right)e^{b\lambda},
\end{equation}
which are found to be:
\begin{eqnarray}
\omega_{\theta}^{\; 12}=1-(1+\alpha f)^{-1/2}, \label{spinconcef}
\end{eqnarray}
the rest being zero or related by symmetry (the spin connection
$\omega$ is antisymmetric in the upper indices \cite{W72}).

The spin connection
\begin{equation}
\Omega_{\mu}=\frac{1}{8}\omega_{\mu}^{\; ab}\left[\gamma_a,\gamma_b\right],
\end{equation}
turns out to be
\begin{eqnarray}
\Omega_{t}=0 ,& \Omega_{r}=0 ,& \Omega_{\theta}= \frac{1-(1+\alpha
f)^{-1/2}}{2}\gamma^{1}\gamma^{2}. \label{spincon}
\end{eqnarray}
 Finally the Dirac equation coupled to the curved surface is:
\begin{equation}
i\gamma^a e_a^{\ \mu}\left(\partial_{\mu}+\Omega_{\mu}\right)\psi=0.
\end{equation}
Substituting all previously computed elements and with some more
algebra, we can cast (\ref{diracc}) into the form:
\begin{equation}
\left[i\gamma^{0}\partial_{0}+i\Gamma(\theta)\partial_{r}+
i\Gamma'(\theta)\frac{\partial_{\theta}}{r}+V(r,\theta)\right]\Psi=0,
\end{equation}
which is the flat Dirac equation in a sort of potential V given
by:
\begin{equation}\label{potential2}
V(r,\theta)=i\Gamma(\theta)\left[1-(1+\alpha
f)^{-1/2}\right](\frac{1}{2r}-\partial_{r}).
\end{equation}

\bibliography{bollo}

\end{document}